\documentclass[useAMS,usenatbib]{mn2e}

\usepackage{graphicx,amssymb,amsmath}

\title[Satellite Galaxy Distribution]{Constraints on the angular distribution of satellite galaxies about spiral hosts}

\author[Steffen and Valenzuela]{Jason H. Steffen$^1$\thanks{jsteffen\_AT\_fnal.gov} and Octavio Valenzuela$^2$ \\
$^1$Fermilab Center for Particle Astrophysics, P.O. Box 500 MS 127, Batavia, IL 60510 \\
$^2$Instituto de Astronom\'ia, Universidad Nacional Aut\'onoma de M\'exico , C.P. 04510, M\'exico, D.F., M\'exico
}

\begin{document}

\maketitle

\begin{abstract}
We present, using a novel technique, a study of the angular distribution of satellite galaxies around a sample of isolated, blue host galaxies selected from the sixth data release of the Sloan Digital Sky Survey.  As a complement to previous studies we subdivide the sample of galaxies into bins of differing inclination and use the systematic differences that would exist between the different bins as the basis for our approach.  We parameterize the cumulative distribution function of satellite galaxies and apply a maximum likelihood, Monte-Carlo technique to determine allowable distributions, which we show as an exclusion plot.  We find that the allowed distributions of the satellites of spiral hosts are very nearly isotropic.  We outline our formalism and our analysis and discuss how this technique may be refined for future studies and future surveys.
\end{abstract}


\section{Introduction}

The standard cosmological model that assumes cold dark matter particles and a cosmological constant as dominant components of our universe (LCDM) has reached a stage where cosmological tests at the scale of galaxies are possible \citep[e.g.][]{avila98, vdBosch98, courteau-rix1999, gnedin06, pizagno2007}.  Although powerful, these types of cosmological constraints are hampered by uncertainties in our understanding of baryonic evolution.  This limitation has led to controversial results \citep[e.g.][]{moore1999, klypin1999} and demonstrate the need for modifications to the model and/or for more detailed comparisons between theory and observations \citep[e.g.][]{valenzuela07, simon-geh07}.

Under the LCDM picture of structure formation, galactic sized halos are assembled by accreting smaller structures or subhalos \citep{kwg1993, ghigna98,klypin1999b}.  Consequently, the presence of satellite galaxies is a natural prediction of hierarchical models and provide important constraints on galaxy-scale dark matter halos that are not always accessible by other means.  For example, it is well known that the radial distribution of satellites can constrain the radial structure of the larger, host dark matter halo \citep{chen06}.  The satellite angular distribution may also provide valuable information because there are indications that the subhalo population is a good tracer of dark matter halo shape \citep{zentner05,agustsson06}.

The pionering observational study on the angular distribution of satellites was presented by \citet{holmberg74}.  This study found an excess of satellites along a direction perpendicular to the host galaxy disk.  The result was later confirmed by \citep{zaritsky97}.  Since the generic LCDM halo is triaxial, a combination of orbital precession and anisotropic dynamical friction might create such a ``Holmberg'' effect \citep{penarrubia04}.  A considerable amount of subsequent work discusses the statistical significance and the observational systematics of the result as well as different interpretations \citep[e.g.][]{bailin07}.

Another prediction of hierarchical models that could contribute to an anisotropic distribution of satellite galaxies is the anisotropic accretion of material along large scale filaments \citep{penarrubia05, zentner05}.  Both the halo shape and its accretion history can imprint a signature on the satellite distribution with respect to the host galaxy and on the orientation of both the host and its satellites with respect to the surrounding large scale structure \citep[e.g][]{aragon07,hahn07,wang07}.

Several recent studies aimed to increase the statistical siginificance of the measurements and to discuss the systematic effects with the hope of constraining the shape of galaxy-scale dark matter halos.  \citet{kroupa05, hartwick00} discuss the case of the Milky Way, \citet{koch06} discuss the M31 system, and \citet{metz2007} disscuss both.  In these cases an anisotropic distribution was found, with relatively high significance, that favors a polar alignment.  Simultaneously, a number of studies adopted the approach of selecting galaxies and their satellites from large surveys and obtain the opposite result; satellites preferentially align with the observed long axis of the host \citep{sl04, brainerd05, yang06, azzaro07, faltenbacher07, kang07}.

\citet{bailin07} recently reviewed the situation and suggested that systematics related to the inclusion of non-relaxed systems (groups or clusters) precluded the detection of the Holmberg effect for late type galaxy hosts in those studies.  The existing discrepancies indicate a need for larger samples, a better understanding and control of systematic effects, and new or complementary analysis approaches in order to have a clear understanding and robust interpretation of the observations and to properly connect the observations to theoretical models.

\subsection{Motivation}

Here we outline a new, complimentary method to study the distribution of satellite galaxies that may serve as a link between numerical work and observations and which may help to resolve some of the outstanding issues in this field.  One challenge that exists in relating numerical studies to observations is centred on the system of coordinates that is used.  From numerical simulations it is relatively straightforward to identify the important physical axes of a galaxy scale dark matter halo.  However, the relative orientation of the dark matter halo to the baryons is not always clear.  In addition, observational studies of satellite galaxies use coordinate systems based upon the profile of the luminous matter as opposed to a system based upon the unobserved dark matter.

This disconnect is particularly relevant if the distribution of satellite galaxies is used as a tracer of the overall shape of the dark matter halo.  Since observations only show a two-dimensional projection of the material, it is not possible to identify which axes of a triaxial halo you are observing and how the system is inclined with respect to the line of sight.  A particular ellipsoid can look either circular or highly elliptical depending upon the location of the observer.  Thus, the observationally motivated coordinate system based upon the distribution of visible matter may be far removed from the physically motivated axes of the dark matter halo; a problem exacerbated by the fact that an LCDM dark matter halo extends far beyond the baryons.  One possible way to circumvent this degeneracy is to use a population of galaxies that have a known, preferred direction that can serve as the basis of a coordinate system that is more readily linked to the physical system of the galaxy and its dark matter halo.

In this article we outline such a method, one which uses disk-like galaxies because the normal to the disk (which can be approximated using the projected galaxy shape) provides the necessary direction of preference, and apply it to data from SDSS DR6, the sixth data release of the Sloan Digital Sky Survey \citep{sdssdr6}.  We also use a parameterization of the angular distribution of satellite galaxies and present our results as an allowed region on an exclusion plot.  This style of presentation extends previous studies which answer the question of whether or not satellite galaxies are distributed isotropically by addressing the question of what distributions are allowed.  Our discussion will proceed with an outline of the formalism that we apply, our data selection criteria, our analysis of those data and its results, and a discussion of the implications of our results, possible ways to improve the sensitivity of this technique, and how this technique can complement other, existing methods used to observationally study dark matter halos.

Before we continue, we note an apparent source of confusion in the literature regarding the coordinate systems used in numerical studies.  In particular, there is a subtle but important difference between the moment of inertia tensor and the tensor of cartesian second mass moments.  For example, the $zz$ component of the moment of inertia tensor \textbf{I} is
\begin{equation}
I_{zz} = \int \rho (x^2 + y^2) dV
\end{equation}
while the $zz$ component of the tensor of cartesian second mass moments \textbf{M} is
\begin{equation}
M_{zz} = \int \rho z^2 dV.
\end{equation}
where $\rho$ is the mass density, and $V$ is the volume occupied by the matter.  We recommend that authors clearly state what quantities they use in their work and take care when using the relevant terminology since the largest moment of inertia corresponds to the smallest cartesian second moment and vice versa.  In the remainder of this work we use the cartesian second moments (or just ``second moments'') exclusively.

\section{Approach and Formalism}

\begin{figure}
\includegraphics[width=0.45\textwidth]{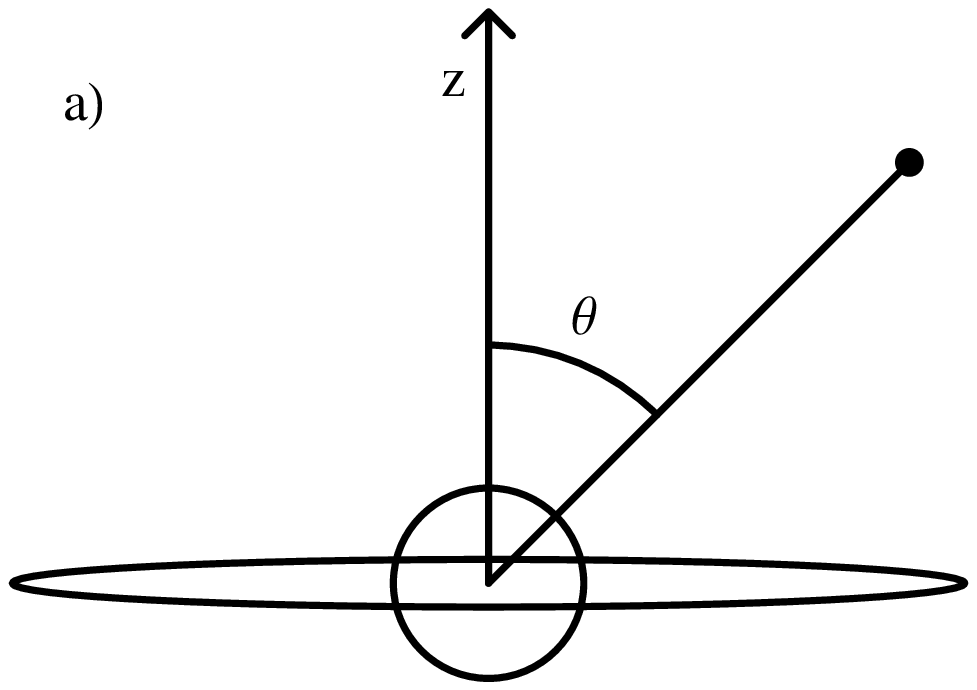}
\includegraphics[width=0.45\textwidth]{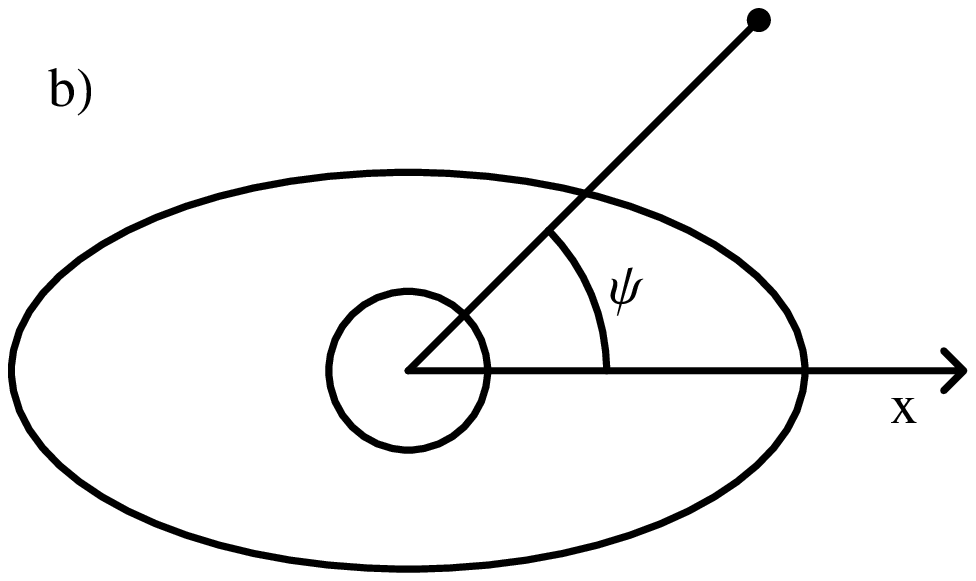}
\caption{a) Diagram of the physical coordinate system constructed about the angular momentum vector of the galaxy.  The angle $\theta$ is the polar angle in a spherical coordinate system.  b) Diagram of the projected coordinate system constructed about the major axis of the observed luminous matter.  The angle $\psi$ is the azimuthal angle in a cylindrical coordinate system where the axis of symmetry is along the line of sight.\label{coordinates}}
\end{figure}

Throughout this article we refer to the ``physical'' and the ``projected'' coordinate systems (see Figure \ref{coordinates}).  The latter system is based upon the distribution of the luminous matter where the primary axis is along the observed major axis of the galaxy.  The former is the coordinate system based upon the angular momentum vector of the galaxy.  Relating this coordinate system to observation requires some assumptions, namely: 1) that the angular momentum vector of a galaxy is normal to the disk (or highly correlated with the normal) and 2) that the inclination of the galaxy can be found using the axis ratio of the isophotal contours.  We will state additional assumptions as they occur.  While not technically an assumption of our analysis, our language will imply that the angular momentum vector of the galaxy is correlated with a principle axis of the dark matter halo (presumably either the largest for a prolate halo or the smallest for an oblate one).

We proceed with the premise that if there is some anisotropy in the distribution of satellites, then the normal to the disk is the natural axis about which to measure the deviations; and that the system can be approximated as cylindrically symmetric.  That is, in the plane of the disk the effects of a triaxial dark matter distribution are negligible or are washed out due to randomness in the orientations of the hosts.  Therefore, a sample of disklike galaxies of various projections allows us to constrain the allowed physical distributions of satellite galaxies.

\begin{figure}
\includegraphics[width=0.45\textwidth]{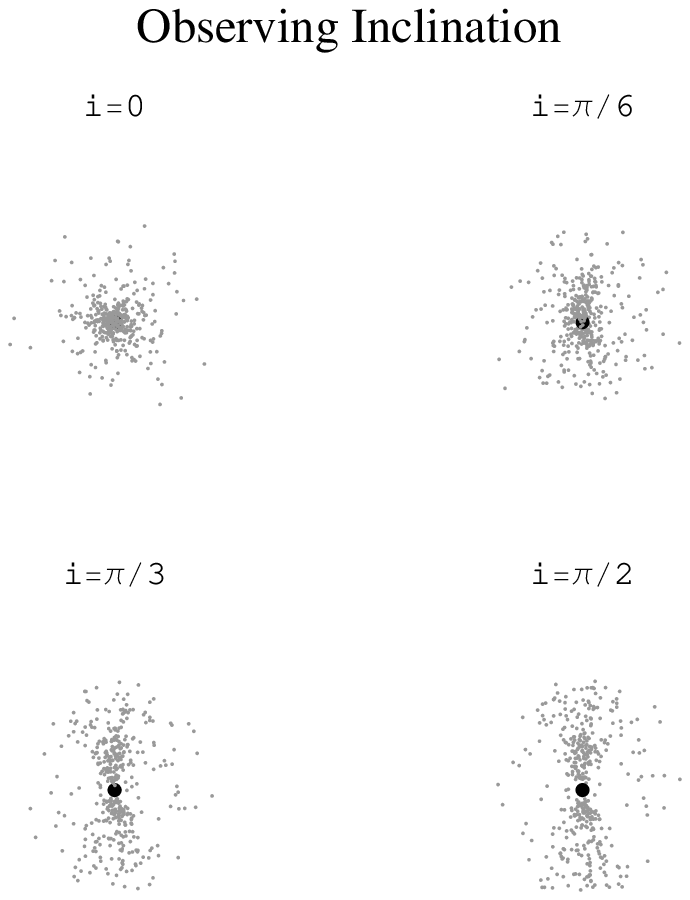}
\includegraphics[width=0.45\textwidth]{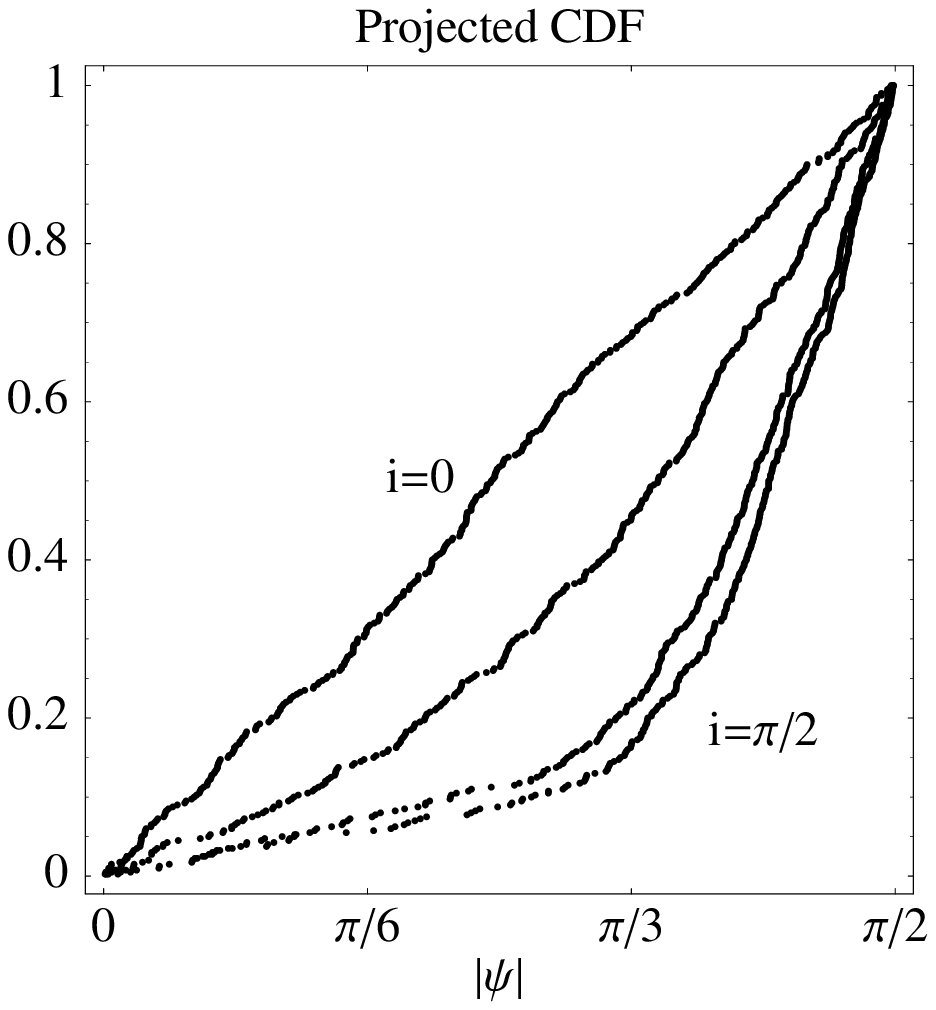}
\caption{Top: Projected distributions of a cylindrically symmetric and highly anisotropic, prolate distribution of satellite galaxies (grey dots) with different inclinations with respect to the line of sight.  Bottom: The cumulative distribution function of satellites for each of the projections.  The value of the angle $\psi$ is the smallest angle between the long axis of the projected disk and the line connecting the centroids of the host and satellite and can have values between 0 and $\pi/2$.  If the distribution were oblate, the CDF would be consistently above the diagonal.\label{rotatepic}}
\end{figure}

For a cylindrically symmetric (though highly anisotropic) distribution of satellites, the projected distribution of satellites is a strong function of the inclination of the galaxy with respect to the line of sight as shown in the top panel of Figure \ref{rotatepic}.  The bottom panel of Figure \ref{rotatepic} shows how the projected cumulative distribution function (CDF) differs from one line of sight to another.  For the projected CDF, the angle $\psi$ represents the minimum angle between the line connecting the centroids of the host and the satellite and the long axis of the projected disk of the host; the absolute value of $\psi$ must be between 0 and $\pi/2$ (see Figure \ref{coordinates}).

In order to study the physical CDF, we first utilize the parameterization of the distribution used by \cite{zentner05}, which is a power law of the form
\begin{equation}
\text{CDF}(x) = a x + (1-a)x^b
\label{abcdf}
\end{equation}
where $x=\vert \cos(\theta)\vert$ and where $\theta$ is the polar angle of the physical coordinate system.  In this parameterization, $a$ characterizes the relative isotropy of the distribution of satellites and $b$ encapsulates some functional form of the deviations from isotropy.  Our analysis approach is designed to constrain the physical distribution of satellite galaxies using a set of projected CDF's, each of which corresponds to a different inclination angle of the galaxy with respect to the line of sight.

\begin{figure}
\includegraphics[width=0.45\textwidth]{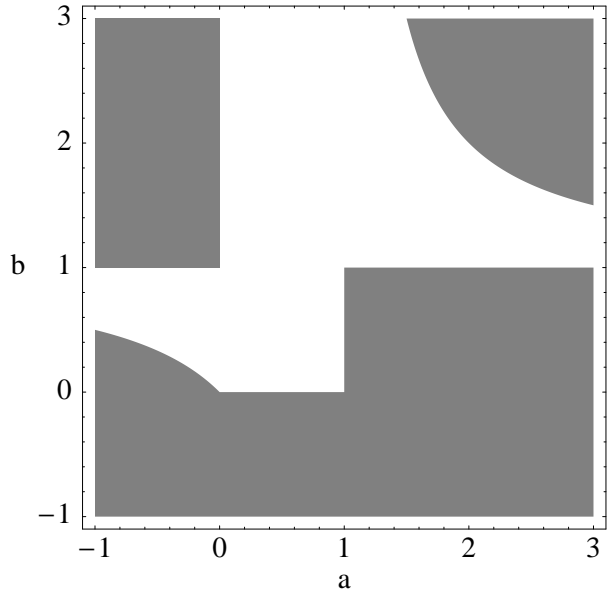}
\caption{Allowed values (unshaded) for the parameters $a$ and $b$ from equation \ref{abcdf}.\label{abfoot}}
\end{figure}

\begin{figure}
\includegraphics[width=0.45\textwidth]{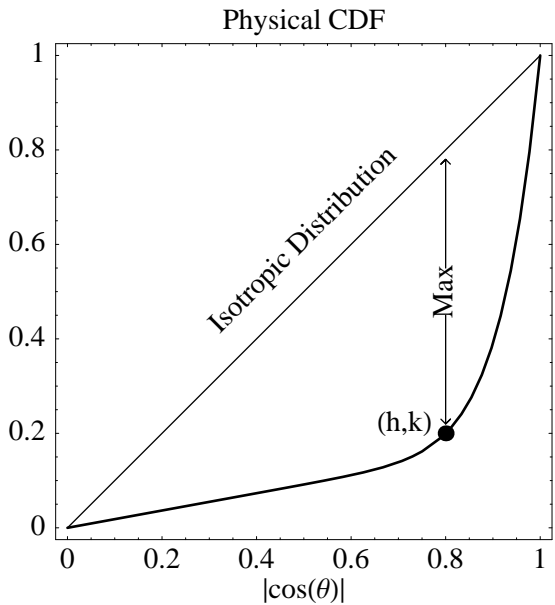}
\caption{Illustration of the relationship between the parameters $a$ and $b$ from equation \ref{abcdf} and the parameters $h$ and $k$.  The point $(h,k)$ corresponds to the location where the physical CDF of the satellite galaxies deviates maximally from the isotropic case.\label{hkplot}}
\end{figure}

By requiring that the physical CDF begin at zero, that it end at unity, and that it has a nonnegative derivative on the interval from 0 to 1, the allowed values for the parameters $a$ and $b$ are restricted to lie in the unshaded region shown in Figure \ref{abfoot}.
While this parameterization has a straightforward functional form, the infinite extent of the footprint--due to the hyperbolic boundary--makes it ill suited for an exclusion plot.  Consequently, we transform the $(a,b)$ parameter space to the $(h,k)$ parameter space where the point $(h,k)$ denotes the point where the physical CDF deviates maximally from the isotropic distribution as shown in Figure \ref{hkplot}.
The transformations between the $(a,b)$ representation and the $(h,k)$ representation are most concisely given by
\begin{equation}
h = b^{1/(1-b)}
\label{abtohk1}
\end{equation}
and
\begin{equation}
a = \frac{k-h^b}{h-h^b}.
\label{abtohk2}
\end{equation}

\begin{figure}
\includegraphics[width=0.45\textwidth]{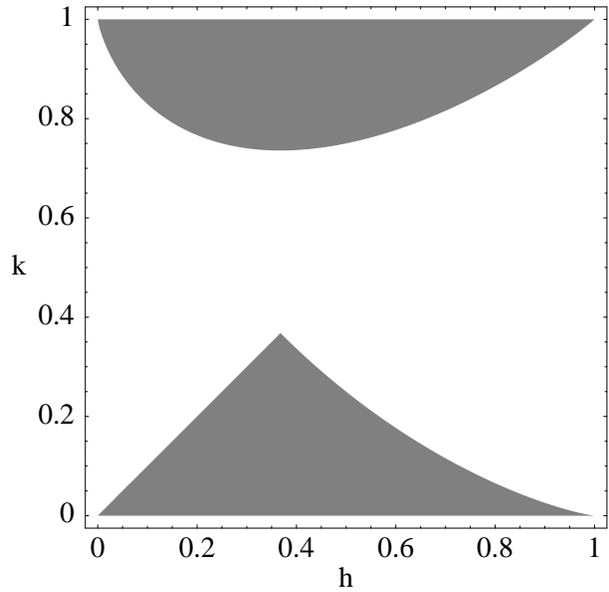}
\caption{Allowed values for the parameters $h$ and $k$ (unshaded) where the point $(h,k)$ corresponds to the location of the point on the physical CDF which deviates maximally from the isotropic case.  This representation allows only finite values for both parameters.\label{hkfoot}}
\end{figure}

There is a degeneracy in this representation that corresponds to the isotropic case (where $b=1$).  In principle this may be an issue, but in practice any problem that arises can be avoided by choosing a representative point for that scenario.  Leaving that item aside, the allowed values of $h$ and $k$ are shown as the unshaded region in Figure \ref{hkfoot}.  Our results will be shown as an allowed region on the $(h,k)$ plane.

\section{Analysis Approach}\label{analysis}

In order to identify the values for $h$ and $k$ that are consistent with data, we divide the $h-k$ plane into a $65\times 100$ grid.   For each point in the allowed parameter space (4290 points) and for each bin we did the following:
\begin{enumerate}
\item Generate a set of satellite galaxies derived from the point $(h,k)$ whose physical distribution is uniform and random in azimuth and uniform but equally spaced in the physical CDF--that is, for each, equally spaced value between 0 and 1 we solve equation (\ref{abcdf}) for $\theta$.  This set is at least 15000 times larger than the number of of satellite galaxies in each bin and serves as the population for our statistical analysis.  The exact number for each bin ($\sim 7 \times 10^6$) is chosen in order to avoid the need to interpolate between points during the analysis.
\item Rotate the population by the midpoint angle of the corresponding bin and rewrite the resulting distribution in the projected coordinate system.
\item From the projected population, draw 1000 samples of galaxies.  The size of these samples are equal to the number of data in each bin.
\item Calculate the sum of the squares of the differences ($\chi^2$) between the data CDF and the population CDF as well as the $\chi^2$ for each of the sample CDF's and the population CDF.
\item Count the number of samples whose $\chi^2$ is smaller than that of the data and assign this value to the point ($h,k$).  We consider this value (divided by 1000) to be the probability that the data were drawn from the distribution associated with ($h,k$).
\end{enumerate}

After completing the above steps for each bin, we make a likelihood function by multiplying the three probabilities for each point $(h,k)$.  Finally, we identify the contours where the likelihood is greater than 0.1 and 0.01 (corresponding to the 90\% and 99\% confidence intervals).

\subsection{Example}\label{example}

Prior to our analysis of SDSS data, we present an example in order to demonstrate the kind of result that this approach yields.  Consider the distribution shown in Figure \ref{rotatepic}.  The parameters that we used to generate those plots are $h=0.8$ and $k=0.15$ ($a=0.115,\ b=12.2$).  We selected a sample of 1200 satellites from that distribution--400 into each of three projection bins with boundaries of $0^{\circ}$, $36^{\circ}$, $54^{\circ}$, and $90^{\circ}$.  Each satellite is selected as though it were viewed from a random orientation within its corresponding bin.  The selection properties and the binning of this sample are chosen to mimic those of the data as explained in the next section.

\begin{figure}
\includegraphics[width=0.45\textwidth]{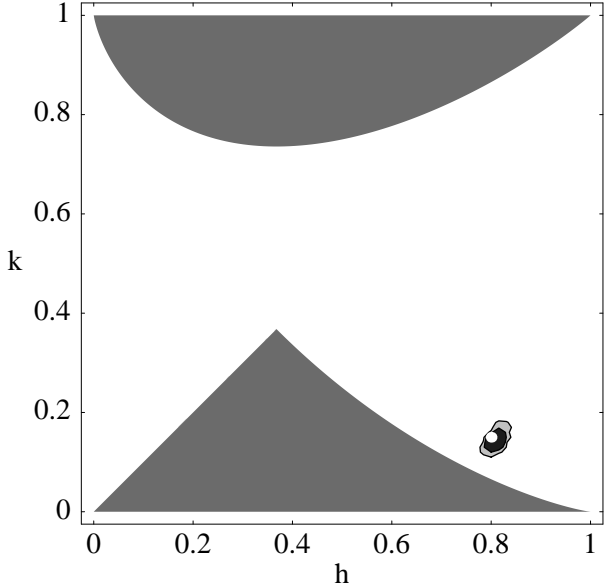}
\caption{Allowed values for $h$ and $k$ for the example shown in Figure \ref{rotatepic}.  The contours correspond to the 90\%, and 99\% confidence levels.  The large gray regions are where the values of $h$ and $k$ do not satisfy the requirements of a cumulative distribution function.\label{contoursample}}
\end{figure}

We apply the analysis approach outlined above to this sample.  The allowed values of the parameters $h$ and $k$ are shown in Figure \ref{contoursample} where the contours correspond to the 90\%, and 99\% confidence levels.  The white dot is the location of the true values of the parameters that were used.

\section{Host and Satellite Sample Selection}\label{sample}

We wish to apply the analysis approach from section \ref{analysis} to a sample of real galaxies.  The galaxies for this study, both hosts and satellites, are selected from the sixth data release of the Sloan Digital Sky Survey spectroscopic and galaxy catalogs \citep{sdssdr6}.  We select galaxies with redshifts between 0.005 and 0.3 and require that their redshifts have a confidence level (\texttt{zconf}) greater than 0.9.  To select our sample of disk-like galaxies, we use the colour divider given in \cite{yang06}
\begin{equation}
^{0.1}(g-r) \leq 0.83.
\label{colorcut}
\end{equation}
This criterion requires that we correct for extinction and that we make k corrections to a redshift of 0.1 (as indicated by the superscript in equation (\ref{colorcut})); for which we use the kcorrect utility \citep{blanton03}.  We include in the sample all galaxies where the axis ratio of the light distribution (using the r-band isophotes) is less than 0.2 since these are likely to be disk galaxies reddened by dust.  We also use the r-band isophotal axis ratio to determine the inclination of each galaxy with respect to the line of sight by assuming that
\begin{equation}\label{inclination}
\cos (i) = B/A
\end{equation}
where $A$ and $B$ are the angular extent of the major and minor axes respectively.

We select our sample of candidate hosts galaxies and satellites following the criteria from \cite{brainerd05}.  That is, a host galaxy must be brighter by one magnitude than any other galaxy: 1) within a projected circular apeture of radius 0.7Mpc and 2) whose relative velocity (calculated using the spectroscopic redshift) is within 1000km/s of the host.  The satellite galaxies are then selected in a similar manner.  A satellite must be two r-band magnitudes dimmer than its host, must lie within a circular apeture of radius 0.5Mpc, and have a relative velocity within 500km/s.  A host/satellites system is rejected if the total brightness of the satellites exceeds that of the host or if there are more than five satellites surrounding the host.  These last criteria are to avoid cases where a single galaxy is deblended into multiple components.  Our final sample includes 1279 host galaxies and 1595 satellites.

We separate these galaxies into three bins of differing axis ratios such that the bins span the entire allowed range of inclinations between 0 and $\pi/2$.  We choose bins such that each of the bins have comparable numbers of satellites.  The first bin is for host galaxies where the angle of inclination between the normal to the disk and the line of sight are between $0^{\circ}$ and $36^{\circ}$--using equation (\ref{inclination}), the second bin has inclinations between $36^{\circ}$ and $54^{\circ}$, and the third bin has inclinations between $54^{\circ}$ and $90^{\circ}$.  These bins contain 385, 514, and 380 host galaxies with 496, 628, and 471 satellite galaxies respectively.  Each bin has approximately 1.25 satellite galaxies per host.

\subsection{Analysis of SDSS Galaxies}

\begin{figure}
\includegraphics[width=0.45\textwidth]{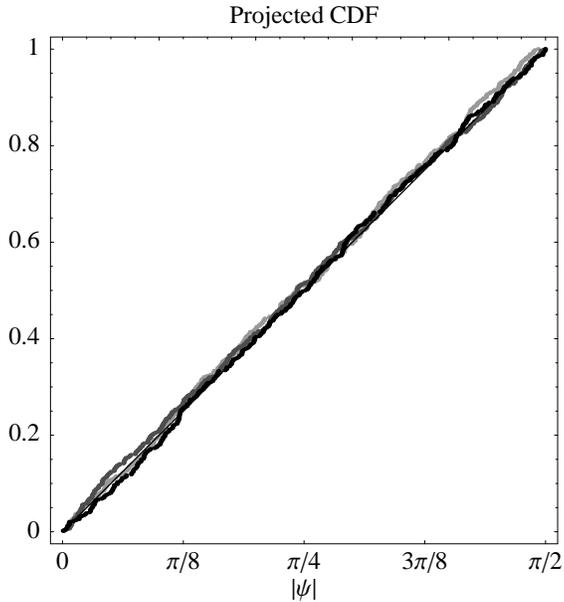}
\caption{The projected CDF's for the different inclination bins.  The light gray points are for the highest inclination bin (nearly edge-on), the dark gray points are for the middle inclination bin, and the black points are for the lowest inclination bin.  The line representing an isotropic distribution in the projected coordinate system is also shown.\label{comparebins}}
\end{figure}

\begin{figure}
\includegraphics[width=0.45\textwidth]{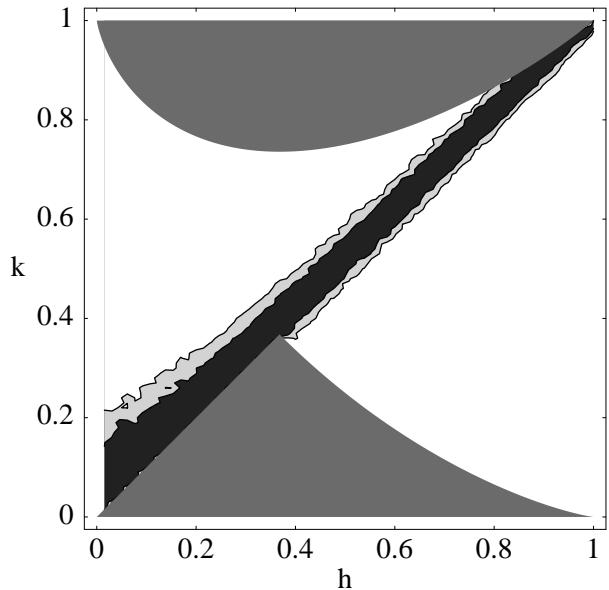}
\caption{Allowed values of $h$ and $k$ for observed, blue galaxies from SDSS DR6.  The contours correspond to the 90\% and 99\% confidence levels.\label{contour}}
\end{figure}

Figure \ref{comparebins} shows the projected CDF's for each of the three inclination bins.  A Kolmogorov-Smirnov test between these different CDF's is inconclusive which indicates that they are consistent with being drawn from the same distribution (which implies that the satellites for this sample are nearly isotropic).
By applying the procedure outlined in section \ref{analysis} to these data we obtain the exclusion plot on the ($h,k$) plane that is shown in Figure \ref{contour}.  This figure shows that the most likely values for the parameters lie along the line $h=k$ and indicates that the physical distribution of satellite galaxies surrounding disk-like hosts is nearly isotropic.
The value of the liklihood function at its maximum is $0.80$ (for the sample in section \ref{sample} it is $0.29$) indicating that the model that we assume for the satellite galaxy distribution (\ref{abcdf}) is consistent with the SDSS data that we analyze.  Moreover, it is unlikely that additional data could demonstrate the inconsistency of the model in the forseeable future.

\section{Discussion}

Certain properties of our galaxy sample, our analysis approach, and our results merit further discussion and as well as comparison with other methods employed to study the distribution of satellite galaxies.  For example, one systematic effect that we did not address is that disk galaxies are not infinitesimally flat, nor are they perfectly axisymmetry when viewed face-on \citep{ryden06}.  The result is that the inclination of edge-on galaxies are underestimated by our approximation in equation (\ref{inclination}) while the inclinations of face-on galaxies are overestimated.

The effect of these issues depends upon the physical distribution of satellite galaxies, whether oblate or prolate, and upon whether the galaxy is edge-on or face-on.  To explore the consequences of this systematic effect, consider four limiting cases: a maximally oblate distribution that is viewed edge-on, the same distribution viewed face-on, a maximally prolate distribution viewed edge-on, and a prolate distribution viewed face-on.

The maximally oblate case is where all the satellite galaxies lie in an annulus in the plane of the disk.  If viewed edge-on, where the inclination is underestimated, the analysis would assume that the apparent line of galaxies was actually drawn from a distribution that was slightly rotated, which in projection would appear elliptical.  Thus, the inferred physical distribution resulting from the analysis of edge-on galaxies is slightly more isotropic (less planar) than what is actually the case.  For face-on galaxies with the same planar distribution, overestimating the inclination would incorrectly assume that the apparent cylindrically symmetric distribution of satellites was again viewed from a slightly different angle.  In this case, the inferred distribution would have, incorrectly, more satellites along the major axis of the host.  The effect of this on the overall inferred physical distribution of satellite galaxies is less straightfoward than for the edge-on galaxies because of the assumption of cylindrical symmetry.  Here the overestimated inclinations give an observed distribution of satellites that is always be anisotropic in the same manner; that is, about the major axis of the observed host.  The result is an incorrect probability assigned to the parameters $(h,k)$ for that projection bin--most of the parameters would have slightly lower probabilities than they should because of the violation of the cylindrical symmetry assumption.  This effect is mitigated somewhat by the fact that the face-on galaxies contribute the least to the final liklihood function.

For the maximally prolate case the satellites would lie in a narrow cone that opens above and below the host disk.  For edge-on galaxies in this scenario, the inferred distribution of satellites would be less prolate (also more isotropic) than the actual distribution.  For face-on galaxies the satellites, as with the oblate case for face-on galaxies, would no longer be cylindrically symmetric but would have more satellites near the observed minor axes of the host (this can be seen in the top panel of figure \ref{rotatepic}, which has different projections of a highly prolate distribution, the edge-on distribution would look less edge-on and the face-on distribution would look less face-on).  Note that the effects for face-on galaxies would be relatively small and contained within the corresponding projection bins.  The more important point of this discussion is that for edge-on systems--which have the highest sensitivity to the physical distribution of satellites--the incorrectly assumed inclination results in an inferred satellite distribution that is more isotropic than what is present in truth.  This is true regardless of whether the distribution is oblate or prolate.

Another issue to explore is the effect of binning the galaxies into either more or fewer bins.  Previous studies effectively used one bin, typically all galaxies with $B/A < 0.8$ (e.g. \cite{brainerd05}).  While in principle it may be possible to identify the physical distribution of satellites by this single, projected distribution (since they are both functions of one variable), there are some systematic effects that this study has brought to light which may complicate that task.  For example, the host galaxies should be randomly oriented with respect to our line of sight.  Yet, the data that we analyze do not have this property.

A likely cause for this discrepancy, particularly for highly inclined galaxies, is dust which affects both the color of a galaxy and its luminosity \citep{shao07}.  Thus, the color cut that we apply does not select a uniform sample as a function of inclination since edge-on galaxies will appear more red.  In addition, extinction causes the edge-on galaxies to appear fainter than their face-on counterparts.  Thus, one would expect fewer edge-on host galaxies in our sample.  This inclination-dependent completeness would need to be addressed if one were using only a single bin of galaxies to infer the physical distribution of satellites.

By increasing the number of inclination bins, one need not assume nor require that the hosts be uniformly sampled in inclination since each bin is largely independent of the others.  Consequently, the results presented here suffer less from this effect than an attempt to identify the satellite distribution from a single bin.  There is no particular limit to the number of inclination bins that one might use to constrain this distribution--we use three for largely computational considerations.  However, once the number of galaxies in each bin becomes too small ($\lesssim 10$) then it would likely be better to alter the approach and use the PDF rather than the CDF, identifying the probability that a particular satellite would be found near a particular angle given the host inclination and the distribution that is then being tested.  Another consideration is that if the difference in inclination between bins is less than the uncertainty in the estimates of the inclination, then there would be a significant amount of overlap between the members of different bins which would complicate the analysis and possibly the interpretation of the results.

\section{Conclusions}

Our results have implications for the distribution of matter in dark matter halos, the orientation of the disk inside of a dark matter halo, and (or) dynamical effects that depend upon the mass of the satellite galaxy.  While these results are consistent with the results of \citet{yang06} for blue host galaxies, there may be some reasons that observed distribution might erroneously imply an isotropic one.  These include: 1) the axis ratios of the disk-hosting dark matter halos might be near unity, resulting in a nearly spherical distribution of satellite galaxies such that our sample cannot distinguish between the actual distribution and a spherical one--something that may be resolved with a larger sample of satellites; 2) the galaxy disks may not be highly correlated with the major or minor axes of the dark matter halo--if the normal to the galactic disk is commonly oriented $\gtrsim 20^{\circ}$ away from the axes of the dark matter halo then the different projection bins would mix and any signal from the dark matter halo would be blurred (see \citet{bailin05, agustsson06}); 3) the selection criteria that we use may not provide an unbiased sample of all of the satellite galaxies in a system--perhaps the dynamics of the relatively bright satellites that we use (ones that would be selected for the SDSS spectroscopic survey) trace the dark matter halo less strongly than smaller galaxies that were not selected.

Future applications of this technique may be able to address some of these issues.  Given the fact that the SDSS data contain millions of galaxies and we only analyze $\sim 1600$, there may be a more sophisticated selection criteria that would identify a larger number of satellite galaxies, especially smaller satellites whose dynamics are more strongly dominated by the host.  In particular, if a method were found that could use photometric redshifts instead of spectroscopic redshifts in the selection criteria, then a much larger sample could be drawn; this would be very useful for data from other existing and planned photometric surveys (e.g. Dark Energy Survey \citep{des}, PanSTARRS \citep{kaiser02}, or LSST \citep{lsst}).  In addition, other selection criteria may eliminate or reduce systematic effects that could cause a biased sample (see \citet{bailin07}).  We chose our selection criteria as a means to tie this technique to previous studies.  Implementing a different or more sophisticated selection criteria lies outside the scope of this work.

If none of the above issues affect our sample then the observed anisotropy reported in \citet{brainerd05} and \citet{yang06} could be explained as the sum of a spherically symmetric distribution around blue galaxies and an anisotropic distribution around red galaxies.  This explanation is consistent with the findings of \citet{yang06}.  The anisotropic distribution of satellites around red galaxies may imply that the merging history plays a major role (if not the primary role) in the triaxiality of dark matter halos and the resulting distribution of satellite galaxies should they prove an unbiased tracer of the mass when the sample is sufficiently large.

Another way to confirm the conclusion that disk-like galaxies reside in nearly spherical halos is by applying this technique to gravitationally lensed systems--both weakly and strongly--where the host galaxy acts as the lens.  While challenging to implement, these probes characterize all of the matter in the lensing system and, in conjunction with the image of the lensing galaxy itself, could help identify the relative orientation of the baryons and the dark matter halo.

\section*{Acknowledgments}

Jason Steffen in supported by the Brinson Postdoctoral Fellowship at the Center for Particle Astrophysics, Fermilab and by the U.S. Department of Energy contract No. DE-AC02-07CH11359.  Octavio Valenzuela thanks support during the initial stage of this project  from the NSF  grant 02-05413  assigned to the Univeristy of Washington and  from the goverment of Mexico through a CONACyT Repatriaci\'on fellowship.  We thank Drs. Eric Agol, Michael Moore, Tereasa Brainerd, and Brian Yanny for useful discussions regarding this work.

\bibliographystyle{mn2e}

\end{document}